\documentclass{article}
\usepackage{preprint}

\usepackage{titling}
\usepackage{lipsum}

\usepackage{authblk}

\usepackage[backend=bibtex,style=numeric-comp, sorting=none]{biblatex}
\usepackage[autostyle=true]{csquotes}
\usepackage{amsmath}
\usepackage{amsfonts}
\usepackage{mathtools}
\usepackage{bm}
\usepackage{booktabs}
\usepackage{silence} 
\newcommand{\mi}{\ensuremath{\mathrm{MI}}}

\newcommand{\sumx}{\ensuremath{\sum\limits_{x \in X} \, }}

\usepackage{dsfont}
\usepackage{braket}

\usepackage[inline]{enumitem}
\usepackage{hyperref}
\usepackage{tabularx}
\usepackage{xfrac}
\usepackage{numprint}
\usepackage{float}
\usepackage[T1]{fontenc}
\usepackage{stfloats}
\usepackage{acronym}
\usepackage{booktabs}
\usepackage{censor}
\usepackage{cleveref}

\usepackage{braket}

\usepackage{epsfig}
\usepackage{subcaption}
\usepackage{calc}
\usepackage{amssymb}
\usepackage{amstext}
\usepackage{amsthm}
\usepackage{multicol}
\usepackage[bottom]{footmisc}

\usepackage{titlesec}
\titlespacing\section{0pt}{12pt plus 3pt minus 3pt}{1pt plus 1pt minus 1pt}
\titlespacing\subsection{0pt}{10pt plus 3pt minus 3pt}{1pt plus 1pt minus 1pt}
\titlespacing\subsubsection{0pt}{8pt plus 3pt minus 3pt}{1pt plus 1pt minus 1pt}

\titleformat{\section}[block]{\centering\Large\bfseries\filcenter}{}{1em}{}
\titleformat{\subsection}[hang]{\bfseries}{}{1em}{}
\titleformat{\subsubsection}[hang]{\bfseries}{}{1em}{}

\pretitle{\begin{center}\huge\bfseries}
\posttitle{\par\end{center}\vskip 0.5em}
\preauthor{\begin{center}\large}
\postauthor{\end{center}}
\predate{\par\large\centering(Dated: }
\postdate{)\par}

\title{Quantum Annealing based Feature Selection in Machine Learning}

\author[1]{Daniel Pranji\'{c}}
\author[1]{Bharadwaj Chowdary Mummaneni}
\author[1]{Christian Tutschku}

\affil[1]{Fraunhofer IAO, Nobelstraße 12, 70569 Stuttgart, Germany}

\date{\today}
\begin{document}

\twocolumn[
	\begin{@twocolumnfalse}

\maketitle
\thispagestyle{empty}

\begin{abstract}
Feature selection is crucial for enhancing the accuracy and efficiency of machine learning (ML) models. This work investigates the utility of quantum annealing for the feature selection process in an ML-pipeline, used for maximizing the mutual information (MI) or conditional mutual information (CMI) of the underlying feature space. Calculating the optimal set of features that maximize the MI or CMI is computationally intractable for large datasets on classical computers, even with approximative methods. This study employs a Mutual Information Quadratic Unconstrained Binary Optimization (MIQUBO) formulation, enabling its solution on a quantum annealer. We demonstrate the capability of this approach to identify the best feature combinations that maximize the MI or CMI. To showcase its real-world applicability, we solve the MIQUBO problem to forecast the prices of used excavators. Our results demonstrate that for datasets with a small MI concentration the MIQUBO approach can provide a significant improvement over MI-only based approaches, dependent on the dimension of the selected feature space.
\end{abstract}
\vspace{0.35cm}
	\end{@twocolumnfalse}]

\section{\uppercase{I. Introduction}}
\label{sec:introduction}

Quantum machine learning (QML) \cite{schuld2019quantum,biamonte2017quantum} is a rapidly evolving field, investigating the intersection of quantum computing and machine learning algorithms. It aims to leverage the unique properties of quantum mechanics, such as superposition and entanglement, to enhance the capabilities of traditional machine learning (ML) methods. For optimal model performance, careful consideration must be given to each stage of the ML pipeline. One crucial step in the pipeline is the so-called feature selection process. Feature selection helps to create simpler models, which in turn reduces computational demands. Utilizing this technique during model training can improve the performance and enhance our understanding of the underlying domain. The high dimensionality of most datasets mandates the development of efficient and effective feature selection algorithms. \\

\noindent Construction machine retailers rely on making well-informed evaluations of residual values. By determining the current and future residual values of their fleet, they can identify the optimal reselling time for individual machines \cite{residualvalue_lucko2007,chiteri2018cash}. Classical ML methods have been extensively applied to calculate the residual value of construction equipment \cite{zong2017maintenance,chiteri2018cash,milovsevic2021estimating,Shehadeh2021,Alshboul2021,stuhler2023benchmarking}. Both conventional ML and automatized ML (AutoML) methods were shown to yield good results for different applications and datasets \cite{Zoller2021a,Zoph2016,stuhler2023benchmarking,horst2024}. In the context of forecasting the residual values of used excavators Support Vector Machines (SVMs) \cite{steinwart} and hybrid (quantum-classical) SVMs have been applied \cite{horst2024,huang,QSVMconcentration}, where quantum SVMs were shown to be competitive with classical SVMs for some encodings. In the latter approach, evaluated quantum kernels serve as input to a constrained quadratic optimization problem which still remians hard to optimize classically. Additionally, quantum methods can be integrated as subroutines in this algorithm, with the expectation of further significant efficiency improvements. E.g., the constrained SVM optimization problem can be reformulated into a quadratic unconstrained binary optimization (QUBO) problem, such that it can be solved on a quantum annealer \cite{finnila1994,date2021qubo}. This approach of solving SVM-QUBOs has lately been studied on annealing devices for real-world problems like biological data classification \cite{date2021qubo,WILLSCH2020107006}. Thus quantum computing can help choose important features in data, opening novel and promising ways to solve complex problems in ML and data analysis. \\

\noindent The study in Ref.~\cite{horst2024} used various feature set combinations from the Caterpillar dataset to rank features based on their importance and impact on model predictions, building upon the work of \cite{voica}. However, it is possible to acquire this knowledge by feature selection algorithms using (un-)supervised methods \cite{jovic2015review}. Especially in the case of unordered categorical features, which need to be encoded in a way that is understood by the ML model, it can be subtle how to weigh all the newly generated features. In this work we demonstrate that the best feature combinations can be obtained for real world problems by using a hybrid approach that uses D-Wave's Kerberos framework \cite{dwave} as a representant of a quantum annealing method for solving QUBOs.\\

\noindent While writing this paper, feature selection in machine learning based on the  mutual information QUBO (MIQUBO) was also addressed in \cite{Hellstern2024}, where its optimization was done differently, using variational quantum algorithms. Their methods were applied to three different datasets \textit{Breast Cancer Wisconsin} \cite{breastcancer}, \textit{Lending Club} \cite{lendingclub} and the \textit{German Credit Data} \cite{germancreditcard}. The latter and most complex dataset includes 20 features (7 numerical and 13 categorical), where one-hot encoding was used for the categorical features. Eventually, they selected 27 out of the 48 one-hot encoded features and 1000 samples, whereas our dataset is larger and higher-dimensional with 2996 samples and  67 features after one-hot-encoding.

\section{\uppercase{II. Methods}}

\begin{figure}[t]
\centering
    \includegraphics[width=\linewidth]{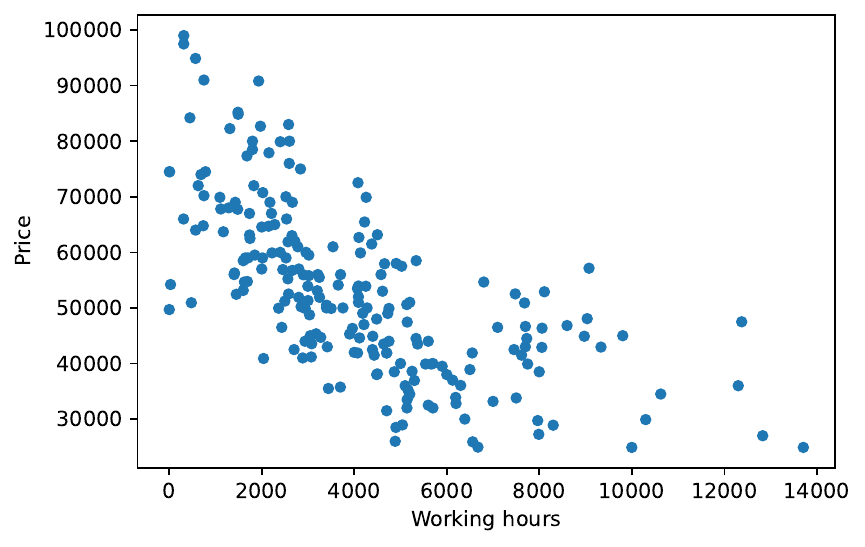}
    \caption{Working hours versus price for the Caterpillar 308 dataset.}
    \label{fig:data-scatter-plot-price-working-hours}
\end{figure}

\begin{table*}[ht]
    \caption{Excerpt of the \textit{Caterpillar (only model 308)} dataset without the extension feature.}
    \label{table:data-excerpt}
    \begin{tabularx}{\textwidth}{X X X X | X}
    \toprule
    CR & Construction~year & Working hours [h] & Location & Price [€]\\
    \midrule
    0 & 2011 & 8973         & FI       & 44.900 €\\
    1 & 2016 & 4183         & PL       & 49.040 €\\
    1 & 2017 & 4655         & PL       & 57.943 € \\
    1 & 2015 & 3175         & GB       & 45.335 € \\
    \bottomrule
    \end{tabularx}
\end{table*}

In the following, we introduce the Caterpillar datasets and the methods used for the quantum annealing based feature selection. \\

\subsection{A. Feature Selection \& Mutual Information}
\label{sec:feature_selection}

\noindent The identification of the most important features of a dataset is an important part of the construction of an ML model to evaluate the prices of construction machines. The dataset \textit{Caterpillar (only model 308)}, has 227 samples and includes categorical features, see Tab.\ref{table:data-excerpt} (we excluded the extension feature). We convert the categories into a binary representation by using one-hot-encoding \cite{bagui2021machine}. This increases the dimensionality from 5 to 27 features. As a reference, we also analyze the larger and more higher-dimensional \textit{Caterpillar (all models)} dataset, which consists of 2996 samples with 6 features out of which 3 are categorical. After one-hot-encoding the dimensionality increases from 6 to 67. These datasets were already analyzed in Ref.~ \cite{horst2024}, where conventional autoencoders were used to obtain a reduced representation that still retains the most relevant information. Although using this technique reduces the dimensionality of the learning problem, the reduced representation is abstract and loses the interpretable connection to the domain. In contrast to the Ref.~\cite{horst2024}, we do not perform any dimensionality reduction schemes in this work, in the sense that the features are being transformed, but only selects the combination of features. However, let us emphasize that in practice, it is common to combine both methods.\\

\noindent A frequently employed and simple feature reduction technique in classical ML is performing a principal component analysis (PCA), where the covariance matrix of the features is diagonalized and the importance of features is then sorted by the size of the resulting eigenvalues $\lambda_i$. In a way, the most important features chosen by this algorithm are those that inhibit the most variance in the dataset. Hence, the explained variance ratio (EVR) is defined as the relative size of $\lambda$ compared to the total variance given as $\mathrm{EVR}\left(\lambda_i \right) = \lambda_i / \sum_{i} \, \lambda_i$. \\

\noindent Besides that, other measures of feature importance than the variance are being used, as well e.g., the so-called mutual information scheme. The mutual information between two random variables $x \in X, y \in Y$ quantifies how much information of $x$ can be obtained from observations of $y$ (and vice versa) and is defined by
\begin{equation}
    \mi(X;Y) = \sum\limits_{x \in X} \, \sum\limits_{y \in Y} \, p(x,y) \log\left(\frac{p(x,y)}{p(x) p(y)}\right) \, , \label{eq:mutual_information}
\end{equation}
where $p(x,y)$ is the joint probability of the marginal probabilites $p(x)$ and $p(y)$. By using the Shannon entropy 
\begin{equation}\label{eq:shannon}
    S(X) = - \sumx p(x) \log p(x) \, ,
\end{equation}
we can rewrite Eq.~\eqref{eq:mutual_information} as 
\begin{equation}
    \mi(X;Y) = S(Y) - S(Y|X) \, 
\end{equation}
where $S(Y|X)$ is defined as: 
\begin{equation}
    S(Y|X) = \sum\limits_{x \in X} \, \sum\limits_{y \in Y} \, p(x,y) \log\left(p(y|x) \right) \, .
    \label{eq:entropy_Y_X}
\end{equation}
Here $p(y|x) $ is the conditional probability of $y$ given $x$ and is defined as $p(y|x) = p(x,y)/p(x) $.  Additionally, we are interested in how much the target $X$ (here: the price) is dependent on a given feature $Y$, given the selection of another feature $Z$. To this end, we can use the conditional mutual information (CMI), defined as
\begin{equation}\label{eq:cmi}
    \mi(X;Y|Z) = S(X|Z) - S(X|Y ; Z) \, 
\end{equation}
where $S(X|Y ; Z)$ is 
\begin{equation}
    S(X|Y ; Z) = \sum\limits_{x \in X} \, \sum\limits_{y \in Y} \, \sum\limits_{z \in Z} \, p(x,y,z) \log\left(\frac{p(x,z|y)}{p(x|y) p(z|y)}\right) \, .
    \label{eq:entropy_Y_X}
\end{equation}
Here, $p(x,z|y) $ is the joint conditional probability of $x$ and $z$ given $y$ and is defined as $p(x,z|y) = p(x,y,z)/p(y) $. Eq.~\eqref{eq:cmi} can be used to determine the most optimal feature combination by selecting the $k$ independent features $X_1, \dots X_k$ of total $n$ features and a target variable $Y$, that maximize the sum of all $\mi(X_i;Y)$ and $\mi(X_j;Y|X_i)$. However, the usability of this approach is limited by the growth of $n \choose k$ in real-world problems. \\

\noindent Instead of calculating Eq.~\eqref{eq:cmi} exactly, several approximations were proposed \cite{venkateswara2015efficient,liu2017fast}. Assuming that there is a conditional independence of features $F_i, F_j$ meaning that $p(F_i | F_j; F_k) = p(F_i | F_k) \, , \forall i,j$ and restricting Eq.~\eqref{eq:cmi} to three features, the optimal combination of features $F$ is approximated by the solution of the following optimization problem
\begin{equation}\label{eq:cmi_approx}
    \underset{F}{\mathrm{argmax}} \, \sum\limits_{i \in F} \left\{ \mi(X_i; Y) + \sum\limits_{j \in F, j \neq i} \, \mi(X_j;Y|X_i) \right\} \, . 
\end{equation}
This is still a hard optimization problem, but it is possible to reformulate it as a QUBO that can be solved on a quantum annealer \cite{mi_effective}.

\subsection{B. Quantum Annealing}
\label{subsec:svm}


\begin{table*}
    \centering
    \caption{One-on-one correspondence of feature selection terms with QUBO problem terms}
    \begin{tabular}{|c|c|c|c|c|}
        \hline
         &Optimization  & Linear Terms & Quadratic Terms& Formula\\
         \hline
        Feature Selection & Maximize & $\mi(X_i;Y)$ & $\mi(X_j;Y | X_i)$ & $\sum\limits_{i \in F} \left \{ \mi(X_i;Y) + \sum\limits_{j \in F, j \ne i} \mi(X_j;Y | X_i) \right \}$\\
        \hline
         QUBO & Minimize & $q_ix_i$ & $q_{i,j} x_i x_j$ & $\sum\limits_{i}^N q_ix_i + \sum\limits_{i<j}^N q_{i,j} x_i x_j$\\
             \hline
    \end{tabular}
    \label{tab:my_label}
\end{table*}

Quantum annealing (QA) \cite{Kadowaki1998} is an optimization process that utilizes quantum fluctuations to find the global minimum of a given objective function over a set of candidate solutions. Unlike simulated annealing (SA) \cite{Kirkpatrick1983}, which relies on thermal fluctuations, quantum annealing (QA) utilizes quantum phenomena such as superposition and quantum tunneling. These quantum effects enable QA to explore a broader and more complex solution space simultaneously, rather than relying solely on random thermal energy fluctuations to escape local minima. Recent studies have compared the performance of QA and SA in terms of computational time for obtaining high-accuracy solutions. Whether QA is superior to SA \cite{Santoro2001, QA_TSP, QIP,Battaglia2005} is still an active area of research. In this work, we contribute to this recent research topic by analysing the performance and suitability of QA for the feature selection process in a ML-pipeline. The development of commercial quantum annealers, such as those based on superconducting flux qubits by D-Wave Systems Inc., have lately led to experimental studies of QA, where the applicability of quantum annealers to several (other) practical problems has been demonstrated \cite{Boyda2017, Neukart2017, Daniel_O}.


\noindent In adiabatic quantum computing (AQC) \cite{McGeoch2014, Albash2018}, the dynamics of a quantum system is described by a time-varying Hamiltonian $H(t)$. The time evolution of the state of a quantum system $\ket{\psi(t)}$ is described by the Schr\"{o}dinger equation
\begin{equation}
    \mathrm{i \hbar} \frac{d\ket{\psi(t)}}{dt} = H(t) \ket{\psi(t)} \, ,
\end{equation}
where $\mathrm{i}$ is the imaginary unit and $\hbar$ is reduced Plancks constant.  
During the time evolution, the system will remain close to the instantaneous ground state of the time-dependent Hamiltonian if the so-called adiabatic condition is fulfilled
\begin{equation}
    \frac{\hbar}{\left(\epsilon_1(t)-\epsilon_0(t)\right)^2} \left|\bra{1(t)} \left|\frac{d H}{dt}\right| \ket{0(t)} \right| \ \ll 1 .
    \label{eq:adiabatic_condition}
\end{equation} 

 Here $\epsilon_0(t)$ and $\epsilon_1(t)$ are eigenvalues of the instantaneous ground state $\ket{0}$ and the excited state $\ket{1}$, respectively. If the Hamiltonian adiabatically transitions from the initial (known) Hamiltonian $H_I$ to the final (unknown) Hamiltonian $H_F$, we can thus find the ground state of an unknown Hamiltonian, by starting in an initialized state. Mathematically, this procedure can be defined by a transition Hamiltonian: 
\begin{equation}
    H = s(t) H_I + (1-s(t)) H_F \, ,
\end{equation}
\noindent where s(t) is a function modeling the transition, such that $s(0) = 1$ and $s(t_F) = 0$ after a certain elapsed time $t_F$. 


 In real-world applications, the ground state of the final Hamiltonian $H_F$, is not just a theoretical construct but a practical tool for solving complex optimization problems. Annealing Systems leverage these principles to solve binary quadratic models (BQM), which include the Ising model \cite{pranjic2019fluctuating,newell1953theory} from statistical mechanics and its computer science counterpart, the quadratic unconstrained binary optimization (QUBO) problem. \\

\noindent In the following, we show the steps to reformulate Eq.~\eqref{eq:cmi_approx} into a QUBO. Given $N$ binary variables $x_1,...,x_N$, the quadratic formulation can be written as 

\begin{equation}
  \min\limits_{x_i, x_j} \left\{ \sum\limits_{i}^N \, q_i x_i + \sum\limits_{i<j}^N \, q_{i,j} x_i x_j \right\} \, , \label{eq:qubo}
\end{equation}
where $q_i$ and $q_{i,j}$ are the linear and quadratic coupling coefficients, respectively. To formulate a problem for the D-Wave system, one must program $q_i$ and $q_{i,j}$, such that the assignments of $x_1, ..., x_N$ represent solutions to the problem. For feature selection, the mutual information QUBO (MIQUBO) method develops a QUBO based on the approximation for $\mi({X_k}; Y)$ (in Eq.~\eqref{eq:cmi_approx}, which can eventually be solved by QA. 

\noindent The restriction that led to Eq.~\eqref{eq:cmi_approx} for MI-based feature selection naturally lends itself to being reformulated as a QUBO. \\

\noindent Each selection of $n \choose k$ features can be represented as the bitstring $x_1, ..., x_N$ by encoding $x_i=1$ if feature $X_i$ should be selected, and $x_i=0$ if not. With solutions encoded in this manner, the QUBO can be represented as $\mathbf{x}^T \mathbf{Q x}$, where $\mathbf Q$ is a $n$ x $n$ matrix and $\mathbf{x}$ is a $n$ x $1$ matrix (a binary vector) that has $k$ ones, representing the $k$ selected features.

\noindent To map Eq.~\eqref{eq:cmi_approx} to a QUBO, we set the elements of $\mathbf Q$ such that $Q_{ii} \mapsto - \mi(X_i;Y)$ and $Q_{i,j} \mapsto - \mi(X_j;Y|X_i)$. These QUBO terms are negative because the quantum computer minimizes the QUBO problem in Eq.~\eqref{eq:qubo}, while the feature selection optimization problem in Eq.~\eqref{eq:cmi_approx} maximizes the CMI.

\begin{figure}
    \centering
    \begin{subfigure}{0.25\textwidth}
        \centering
        \includegraphics[width=0.5\linewidth]{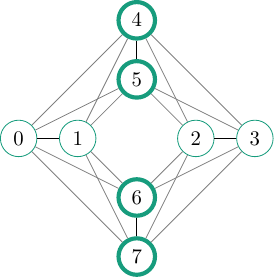}
        \caption{}
    \end{subfigure}%
   \begin{subfigure}{0.25\textwidth}
        \centering
        \includegraphics[width=0.8\linewidth]{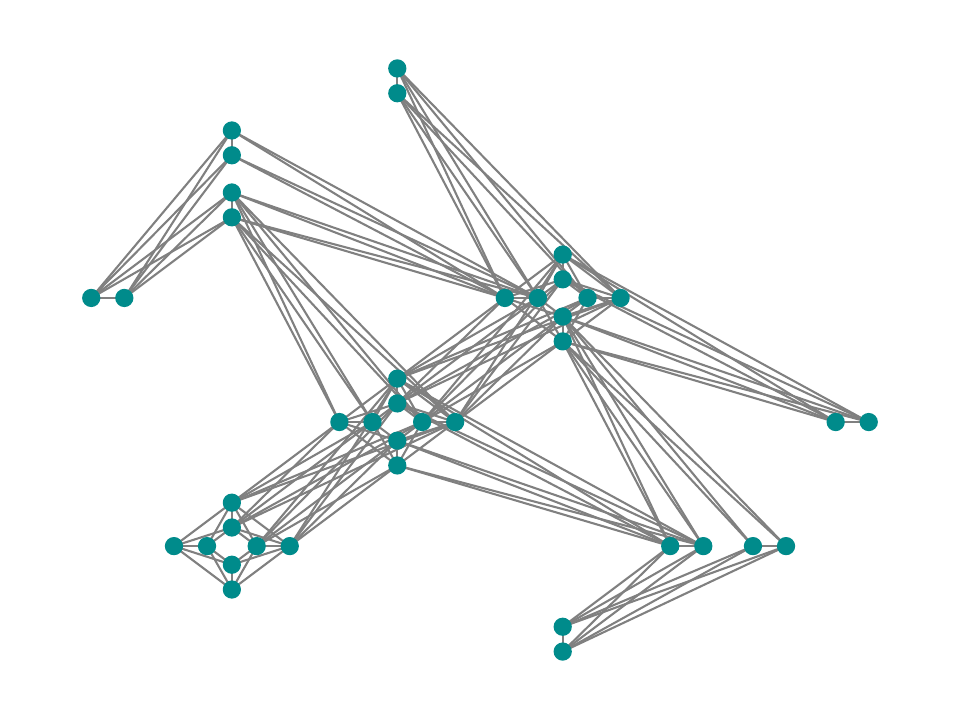}
        \caption{}
    \end{subfigure}
    \caption{(a) Single cell of the Pegasus architecture, (b) repeating single cells to form a unit cell.}
    \label{fig:Pegasus architecture}
\end{figure}
\begin{figure}
    \centering
    \includegraphics[width=0.5\linewidth]{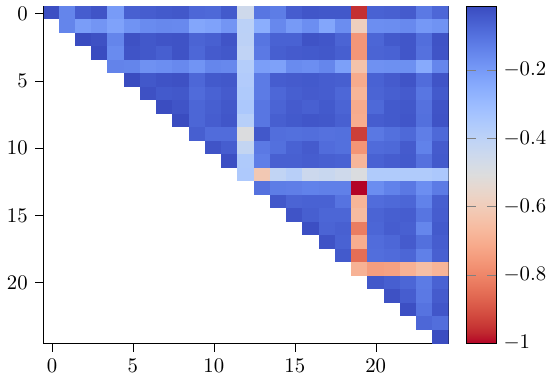}
        \caption{Sparsity pattern of the feature selection problem, formulated as QUBO. Only upper triangular matrix is shown.}
    \label{fig:Sparsity plot}
\end{figure}

The D-Wave hybrid sampler, Kerberos, tackles large and intricate problems by employing a blend of classical and quantum approaches. It simultaneously runs three algorithms in parallel: Tabu search, simulated annealing, and QPU sub-problem sampling. Tabu search is a heuristic optimization technique that incorporates local search strategies and is designed to avoid getting trapped in local optima by maintaining a list of previously visited solutions, known as the tabu list. The duration for which solutions remain in this list is referred to as the tenure. The dwave-samplers package includes an implementation of the MST2 multi-start tabu search algorithm \cite{Palubeckis2004}, which is specifically designed for solving QUBO problems. SA is used to escape the local minima and find an approximate global minimum.  Each of the three routines thus tackles specific aspects of the problem, and Kerberos intelligently combines their best results to deliver superior solutions. Kerberos employs clique embedding, a technique that identifies highly interconnected problem subsets (cliques) that significantly influence the solution. These cliques are then fed to the QPU for focused analysis. This approach maximizes the efficiency of quantum resources by directing them towards critical areas and avoids wasting them on less impactful parts. In this work, for a larger dataset after keeping the important features $(25)$ with the help of mutual information, $95$ physical qubits were used for QPU sampling. The features get embedded onto the physical qubits on the Dwave advantage system with the pegasus architecture in Fig. \ref{fig:Pegasus architecture}. To embed each feature, the maximum number of physical qubits required is $4$. The sparsity of the associated QUBO matrix of the given problem is shown in Fig. \ref{fig:Sparsity plot}. A quick glance at the elements in the QUBO matrix provides some insight into the seemingly most important features. The selected features correspond to the lowest energy states, aligning with optimal solutions, and have high measurement probabilities, ensuring consistent selection. This combination allows the algorithm to reliably target the best solutions for the problem.

\begin{table}
    \centering
    \begin{tabular}{|c|c|}
        \hline
        Sampling time & 286.128 $\mu$s\\
        \hline
        Anneal time per sample& 600 $\mu$s\\
         \hline
         Readout time per sample& 95 $\mu$s\\
        \hline
    \end{tabular}
    \vspace{3mm}
    \caption{Annealing schedule information}
    \label{tab:my_label}
\end{table}

\subsection{C. Support Vector Regression}

SVMs represent a prominent class of machine learning algorithms adept at both linear and non-linear classification and regression tasks. They are particularly excelling with small to medium-sized datasets. For regression tasks, the SVM learns an approximator that maps the input to a continuous output. For this, the model minimizes the error margin between the predicted and target values. Data points closest to this margin (called support vectors) play a crucial role in shaping the approximator and influencing the model's behavior. \\

\noindent By applying a transformation called the \textit{feature map} $\phi$, we can project the data into a higher-dimensional space, enabling models to capture complex, non-linear relationships between features and the target variable. However, this $\phi$ incurs a significant computational expense, when the coordinates have to be computed for each data in that feature space. The kernel trick offers a computationally efficient alternative, where the feature space is utilized implicitly by leveraging kernel functions, for example the radial basis function (rbf), given by $K(x,y) = \exp(-\gamma |x-y|^2)\, , \gamma>0$. The kernel can be interpreted as a measure of similarity between data points in that feature space. This results in a kernel matrix that effectively captures the relationships within the transformed space. \\

\noindent We'll proceed by introducing the formulation of SVMs for regression and explain how to obtain the support vectors in the following. Given a set of samples $\left\{ (x_1, z_1), \dots (x_l, z_l) \right\}$ with input data $x_i \in \mathbb{R}^n$ and target output $z_i \in \mathbb{R}$, the primal form of the optimization problem is given by
\begin{equation}\label{eq:svm}
    \begin{split}
        \min\limits_{w \in \mathbb{R}^n, b \in \mathbb{R}, \xi, \xi^*} \quad &\frac{1}{2} \left|\left| w \right|\right|^2 + C \sum\limits_{i=1}^l \, \xi_i + C \sum\limits_{i=1}^l \, \xi^*_i \, , \\
        \text{subject to} \quad &w^T \phi(x_i) + b - z_i \le \varepsilon + \xi_i \, , \\
        &z_i - w^T \phi(x_i) - \le \varepsilon + \xi^*_i \, , \\
        &\xi_i, \xi^*_i \ge 0 \, , i = 1, \dots, l \, ,
    \end{split}
\end{equation}
where $\epsilon > 0$ (insensitivity of the loss \cite{steinwart}) and $C > 0$ is a regularization parameter. Because $w$ can be very high-dimensional, in practice the dual problem is solved instead
\begin{equation}\label{eq:dual_svm}
    \begin{split}
        \min\limits_{\alpha, \alpha^*} \quad &\frac{1}{2}(\alpha - \alpha^*) K (\alpha - \alpha^*) + \varepsilon \sum\limits_{i=1}^{l} \, (\alpha_i + \alpha^*_i) \\ 
        &+ \sum\limits_{i=1}^l \, z_i (\alpha_i - \alpha^*_i) \, ,\\
        \text{subject to} \quad &\sum\limits_{i=1}^l \, (\alpha_i - \alpha^*_i) = 0 \, , \\ 
        &0 \le \alpha_i, \alpha^*_i \le C \, , i=1, \dots, l \, ,
    \end{split}
\end{equation}
where $K_{ij} = K(x_i, x_j) = \phi(x_i)^T \phi(x_j)$ is called the \textit{kernel} and $\alpha_i, \alpha^*_i$ are called support vectors. Details on the computational aspects of solving Eq.~\eqref{eq:dual_svm} can be found in Ref.~\cite{chang2011libsvm}. After obtaining the optimal $\alpha_i, \alpha^*_i$, the approximator $f$ is given by

\begin{equation}\label{eq:approximator}
    f(x) = \sum\limits_{i=1}^l \, (-\alpha_i + \alpha^*_i) K(x_i, x) + b \, .
\end{equation}
This SVM formulation for regression can be extended with a parameter $\nu \in (0,1]$ that controls the ratio of support vectors used in the model and the error margin \cite{scholkopf2000}. This can be helpful especially for real-world datasets where a clear margin might be hard to define \cite{scholkopf2000,suganyadevi2014support}. \\

\noindent In our models we evaluate the performance using the so-called $\mathrm{R2}$ score, defined by
\begin{equation}\label{eq:r2_score}
    \mathrm{R2} = 1 - \frac{\sum\limits_{i=1}^n \, (z_i - \hat{z}_i)^2}{\sum\limits_{i=1}^n \, (z_i - \bar{z}_i)^2} \, ,
\end{equation}
where $z_i, \hat{z}_i, \bar{z}_i$ are the true, predicted and average target outputs respectively. If $z_i = \hat{z}_i \, , \forall i$, or in other words, a perfect prediction, then $\mathrm{R2}=1$. 

\begin{figure*}
    \centering
	\begin{subfigure}{0.45\textwidth}
    \caption{}\label{fig:mi_miqubo_a}
		\includegraphics[width=\textwidth]{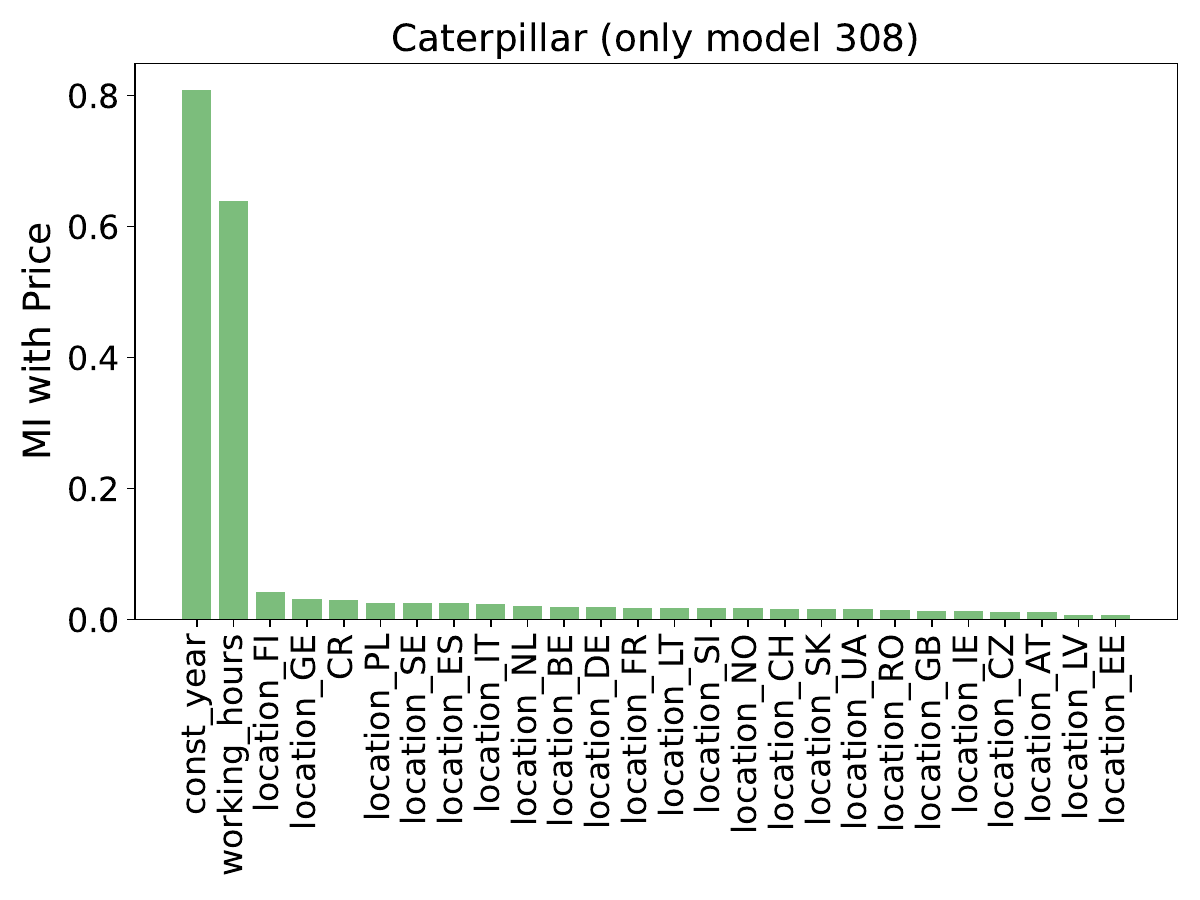}
		
	\end{subfigure}
 	\begin{subfigure}{0.45\textwidth}
            \caption{}\label{fig:mi_miqubo_b}
		\includegraphics[width=\textwidth]{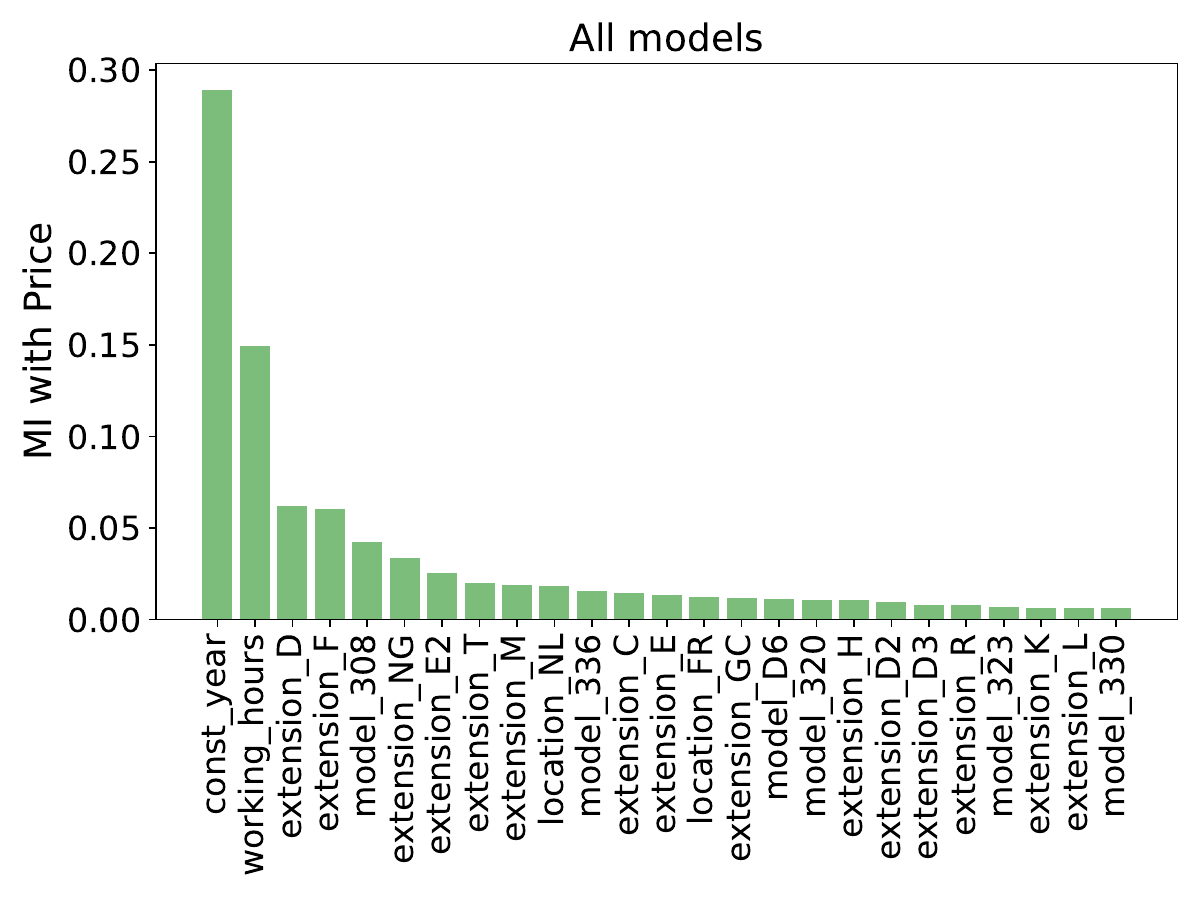}
		
	\end{subfigure}
 	\begin{subfigure}{0.45\textwidth}
            \caption{}\label{fig:mi_miqubo_c}
		\includegraphics[width=\textwidth]{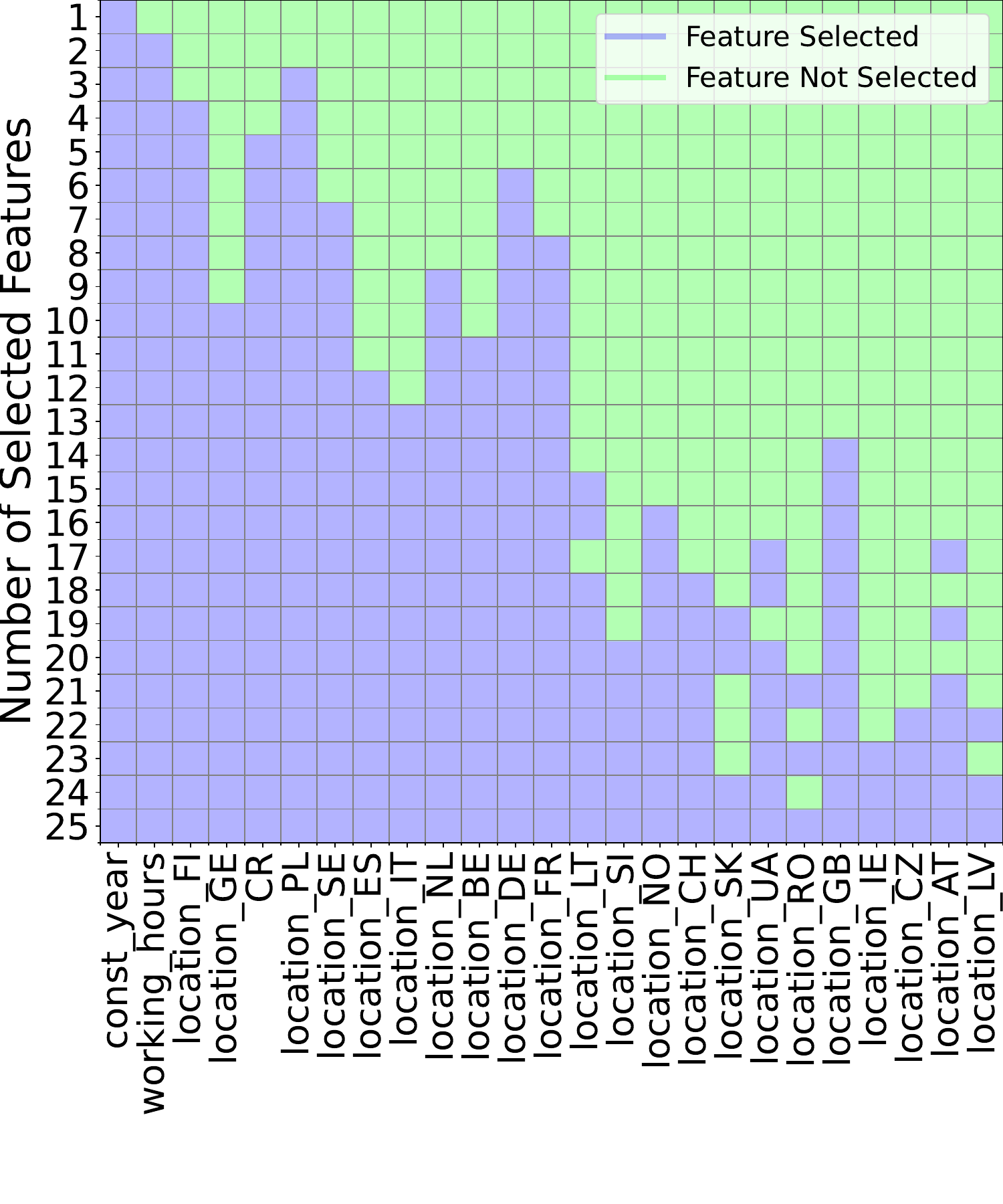}
		
	\end{subfigure}
 	\begin{subfigure}{0.45\linewidth}
            \caption{}\label{fig:mi_miqubo_d}
		\includegraphics[width=\linewidth]{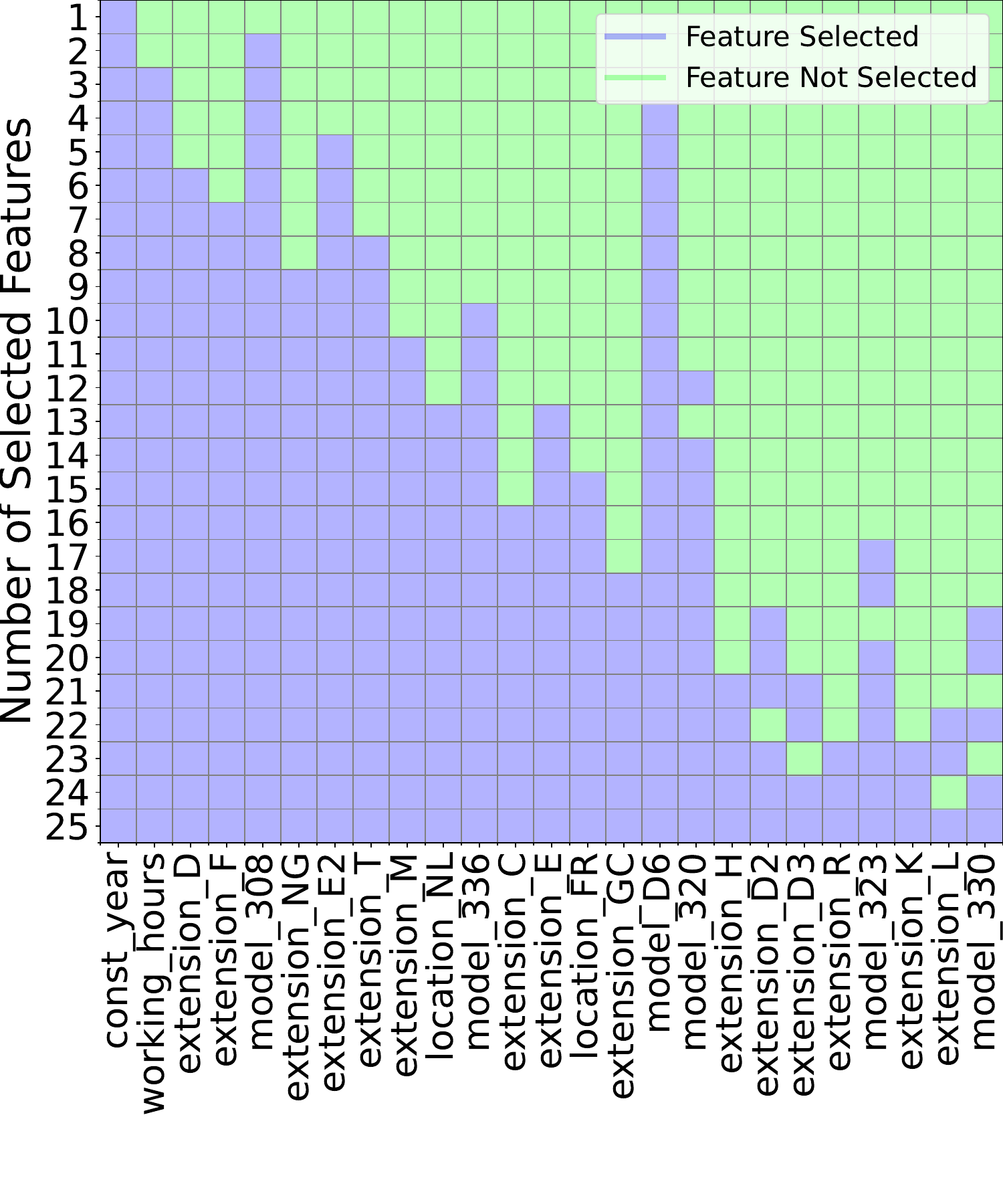}
		
	\end{subfigure}
    \caption{One-hot-encoded features of the a) \textit{Caterpillar (only model 308)} \& b) \textit{Caterpillar (all models)} dataset are shown with the most mutual information (MI) between an individual input feature and the target output (here: price). a) shows that the dataset is MI-concentrated as most of the MI localized in a small subset of all features, whereas in b) it is more equally spread over the features, hence less MI-concentrated. c) \& d) shows the combination of features that maximizes Eq.~\eqref{eq:cmi_approx} for a given number $k$ of selected features for the c) \textit{Caterpillar (only model 308)} \& d) \textit{Caterpillar (all models)} datasets. The deviation from the diagonal in c) \& d) indicates that the set of features that maximizes the CMI in Eq.~\eqref{eq:cmi_approx} does not match an MI-based selection.}
    \label{fig:mi_miqubo}
\end{figure*}

\section{\uppercase{III. Results}}
\label{sec:results}

Fig.~\ref{fig:mi_miqubo_a} shows the one-hot-encoded features with the highest MI of the \textit{Caterpillar (only model 308)} dataset. The result reflects that the construction year and the working hours of a used excavator are the features that carry the most mutual information about its price. This is further indicated by the fact that there is a quick drop in the mutual information from approximately 0.8 \& 0.6 to less than 0.1. We refer to this as an example of a MI-concentrated dataset. The information contained in the \textit{location} feature is distributed into many individual features $\left\{ \textit{location}\_\textit{FI}, \textit{location}\_\textit{PL}, \dots \right\}$, due to one-hot-encoding. In contrast, for the \textit{Caterpillar (all models)} dataset, the MI is more equally distributed over the features and is hence a less MI-concentrated dataset cf Fig.~\ref{fig:mi_miqubo_b}. A less MI-concentrated dataset is expected to have more deviations in the selection of the feature combination that maximizes the conditional mutual information, compared to the MI-only feature selection.

\begin{figure*}
    \centering
    \includegraphics[width=\linewidth]{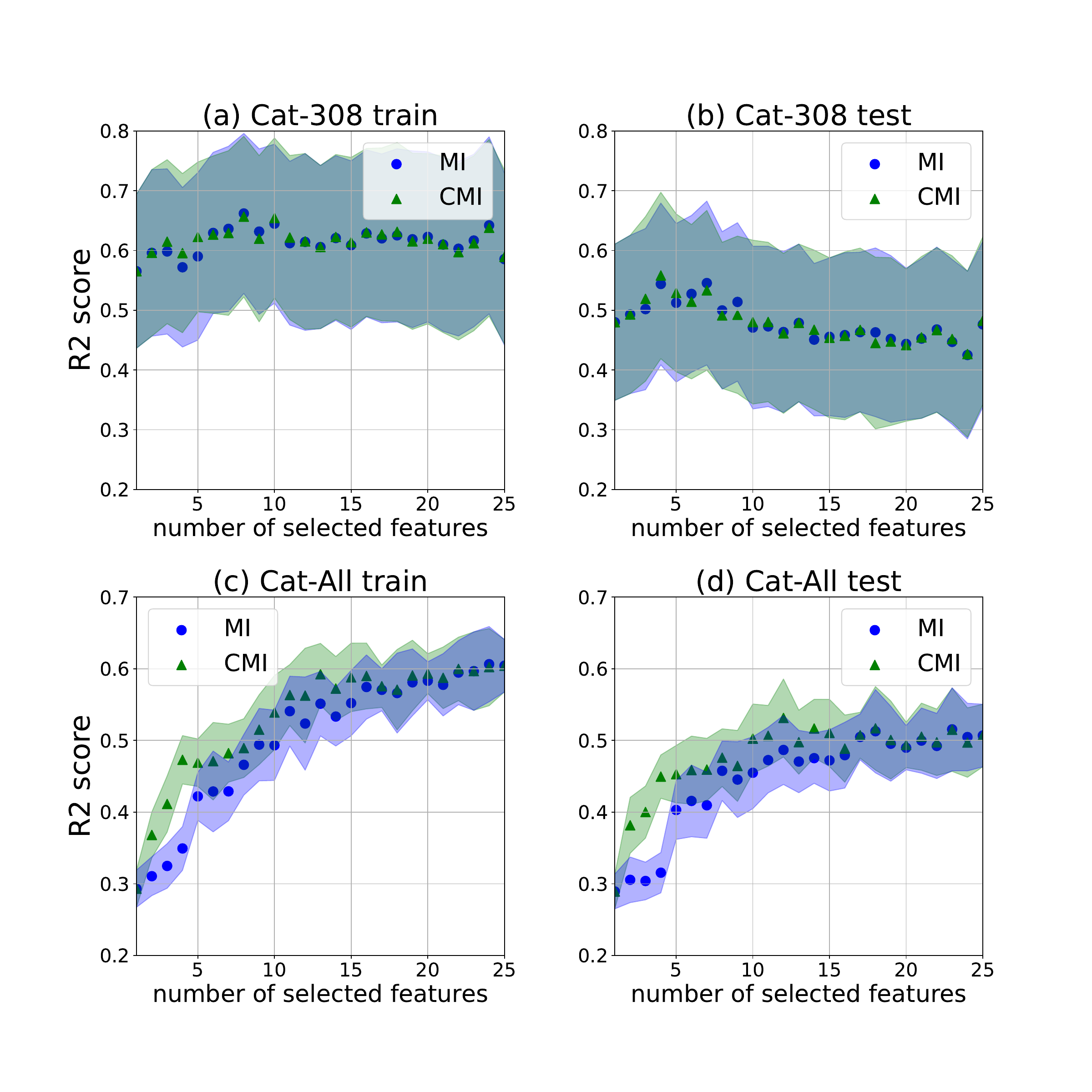}
    \caption{Mean $\mathrm{R2}$ scores of the rbf-SVMs evaluated on several splits of \textit{Caterpillar (only model 308)} \& \textit{Caterpillar (all models)} datasets, denoted by Cat-308 \& Cat-All. For Cat-308 we used 100 randomly generated train-test-splits, whereas for Cat-All we used 15. The $k$ features with the highest mutual information are selected for 'MI' and the $k$ features that maximize the conditional mutual information are denoted by 'CMI'.
    (a) \& (b) For the MI-concentrated Cat-308 dataset, there is no significant difference in the R2 scores observed. (c) \& (d) For the less MI-concentrated Cat-All dataset, there is a drastic performance gap in the performance for a lower number of selected features when maximizing the sum of MI \& CMI instead of just MI. For a higher number of selected features, the difference in information between the selected features decreases and hence doubled for both datasets, the models converge towards the same mean $\mathrm{R2}$ score. Parameters: $\gamma = C = 1, \varepsilon = 10^{-3}$.}
    \label{fig:result_SVM}
\end{figure*}

\noindent Fig.~\ref{fig:mi_miqubo_c} and Fig.~\ref{fig:mi_miqubo_d} show for a given number $k$ the best combination of features that maximizes the combined MI and CMI among the selected features. The two features \textit{const\_year} \& \textit{working\_hours} were chosen consistently through majority of the selected features $k$. On the contrary side in Fig. \ref{fig:mi_miqubo_a} \textit{location}\_\textit{GE} (Georgia) has more MI with the price than \textit{location}\_\textit{DE} (Germany), but the set of features that maximizes the CMI for $k=6$ includes \textit{location}\_\textit{DE} instead of \textit{location}\_\textit{GE}. Although \textit{location}\_\textit{DE} is not in the top 6 features measured by its MI with the price, it is one of the top 6 of the chosen features by CMI.  \\

\noindent In the following, we demonstrate empirically that a set of $k$ selected features with maximized the sum of MI and CMI performs better in a typical machine learning model, than one that maximizes only the total MI instead. From an information theory perspective, the reason for that is that the former set of features contain more independent information than the latter \cite{ash2012information}. In Fig.~\ref{fig:result_SVM} we evaluated the mean R2 score of a rbf-SVM model for 100 train-test-splits on the \textit{Caterpillar (only model 308)} (see Fig. \ref{fig:result_SVM}a and \ref{fig:result_SVM}b) and 15 train-test-splits on the \textit{Caterpillar (all models)} dataset (see Fig. \ref{fig:result_SVM}c and \ref{fig:result_SVM}d). The data in both cases was standardized and centralized before the kernel evaluation, while we fixed $\gamma = C = 1$ and $\varepsilon = 10^{-3}$. \\

\noindent For Fig.~\ref{fig:result_SVM}a and ~\ref{fig:result_SVM}b there is no visible difference between the method of maximizing MI in comparison to maximizing the CMI. This is due to the fact that \textit{Caterpillar (only model 308)} is a dataset with more MI-concentration than \textit{Caterpillar (all models)}. Hence in Fig.~\ref{fig:result_SVM}c and \ref{fig:result_SVM}d, where the MI-concentration is much less, a statistically relevant gap according to the standard deviation between both feature selection methods opens. The difference becomes most evident for $k \le 5$. Eventually, the difference between the information content for the set selected by MI and CMI vanishes as both methods include most of the highly informative features, reducing the performance gap between them.

\section{\uppercase{IV. Conclusion}}
\label{sec:conclusion}

This study showcased the usefulness of quantum annealers for feature selection in machine learning. The challenge of finding the optimal feature set for large datasets was addressed by leveraging the formulation of MI Quadratic Unconstrained Binary Optimization (MIQUBO). A hybrid annealing approach was then used to solve the MIQUBO. To demonstrate the usefulness of MIQUBO in identifying the best features, we applied the obtained combinations to a real-world scenario: predicting excavator prices. In particular, we demonstrate that for datasets with less MI-concentration the set of features that maximizes the sum of MI and CMI leads to a better performance of machine learning models for small latent subspaces. Further research is needed to investigate the application of MIQUBO to different datasets and machine learning tasks, exploring the broader potential of quantum computing in enhancing the performance and efficiency of machine learning models. It would be interesting to identify the performance gap in other datasets that have even less MI-concentration, such as in more complicated data like images, audio etc. It is worth noting that the QUBO-matrices that arise in our problem are relatively sparse and would be more dense, if there is less MI-concentration in the data, which is mandatory for quantum advantage.

\section*{ACKNOWLEDGEMENTS} 
This work was funded by the German Federal Ministry of Economic Affairs and Climate Action in the research project AutoQML (grant no. 01MQ22002A). The authors thank Horst St\"{u}hler for providing the Caterpillar datasets.

\section*{BIBLIOGRAPHY}

%
\printbibliography[heading=bibintoc]

\end{document}